\documentclass[aps,prb,twocolumn,groupedaddress]{revtex4-1}
\usepackage[latin9]{inputenc}
\usepackage{graphicx}
\usepackage{epstopdf}
\usepackage{xcolor}
\makeatletter
 
 \@ifundefined{textcolor}{}
 {%
   \definecolor{BLACK}{gray}{0}
   \definecolor{WHITE}{gray}{1}
   \definecolor{RED}{rgb}{1,0,0}
   \definecolor{GREEN}{rgb}{0,1,0}
   \definecolor{BLUE}{rgb}{0,0,1}
   \definecolor{CYAN}{cmyk}{1,0,0,0}
   \definecolor{MAGENTA}{cmyk}{0,1,0,0}
   \definecolor{YELLOW}{cmyk}{0,0,1,0}
 }

\usepackage{dcolumn}
\usepackage{bm}

\makeatletter
\newcommand{\colorcaption}[2][]{%
  \begingroup%
  \renewcommand{\@caption@fignum@sep}{ (color online). }%
  \caption[#1]{#2}%
  \endgroup%
}
\makeatother

\makeatother

\begin{document}






\title{Computational study of the shift of the G band of double-walled carbon nanotubes due to interlayer interactions}

\author{Valentin N. Popov}
\affiliation{Faculty of Physics, University of Sofia, BG-1164 Sofia, Bulgaria}

\author{Dmitry I. Levshov}
\affiliation{Laboratoire Charles Coulomb (UMR5221), CNRS-Universit\'{e} de Montpellier, F-34095 Montpellier, France}
\affiliation{Faculty of Physics, Southern Federal University, 344090, Rostov-on-Don, Russia}

\author{Jean-Louis Sauvajol}
\affiliation{Laboratoire Charles Coulomb (UMR5221), CNRS-Universit\'{e} de Montpellier, F-34095 Montpellier, France}

\author{Matthieu Paillet}
\affiliation{Laboratoire Charles Coulomb (UMR5221), CNRS-Universit\'{e} de Montpellier, F-34095 Montpellier, France}

\date{\today}
\begin{abstract}
The interactions between the layers of double-walled carbon nanotubes induce measurable shift of the G bands relative to the isolated layers. While experimental data on this shift in free-standing double-walled carbon nanotubes has been reported in the past several years, comprehensive theoretical description of the observed shift is still lacking. The prediction of this shift is important for supporting the assignment of the measured double-walled nanotubes to particular nanotube types. Here, we report a computational study of the G-band shift as a function of the semiconducting inner layer radius and interlayer separation. We find that with increasing interlayer separation, the G band shift decreases, passes through zero and becomes negative, and further increases in absolute value for the wide range of considered inner layer radii. The theoretical predictions are shown to agree with the available experimental data within the experimental uncertainty.     
\end{abstract}
\maketitle


\section{Introduction}

In the last few decades, the sp${^2}$ carbon-based materials like fullerenes, nanotubes, graphene and few-layer graphene have been a subject of intense experimental and theoretical investigations because of their unique properties, originating from their zero-, one-, two-, and three-dimensionality.\cite{jori11} In particular, significant progress has already been achieved in the synthesis and the study of the properties and application of carbon nanotubes in bulk samples as well as at the single nanotube level.\cite{rtm04,jori08,jari13,liu17,fuji16,devo13,sait11,levs17} The application of nanotubes in nanoelectronics requires their precise structural characterization. For this purpose, the Raman scattering of light by phonons is the experimental technique of choice, being a fast and nondestructive characterization method.\cite{thom07} 

The single-walled nanotube (SWNT) can be viewed as obtained by wrapping up of a graphene sheet into a seamless cylinder. It has a few intense Raman bands, arising from the radial-breathing mode (RBM) and the totally symmetric longitudinal and transverse G modes (also named tangential G modes). In semiconducting SWNT, the longitudinal G mode has higher frequency than the transverse one and vice versa in metallic ones. However, the observed higher-frequency G band is always denoted as G$^+$ and the lower-frequency one as G$^-$. The RBM frequency is found to be inversely-proportional to the nanotube radius and is normally used for fast sample characterization.\cite{arau08} The G modes also depend on the nanotube radius and can be used for supporting the assignment of the spectrum to particular nanotube but contain additional information that allows differentiating between metallic and semiconducting nanotubes.\cite{jori02,mich09,pail10} 

The double-walled carbon nanotube (DWNT) is a layered structure, consisting of two nested SWNTs bound together by weak Van der Waals interactions. The Raman spectra of $C_{60}$-derived DWNTs are found to exhibit several intense bands due to radial-breathing like modes (RBLMs) and G modes of the two layers.\cite{pfei08,villa10} The modification of the RBLMs by interlayer interactions can be modeled straightforwardly within continuum models\cite{popo02,doba03} and the derived results can be applied directly to the assignment of the RBLMs of DWNTs.\cite{pfei04} It has also been observed that the G bands of the inner $(6,5)$  layers of $C_{60}$-derived DWNTs shift with respect to those of the isolated layers due to interlayer interactions.\cite{villa10} G-band shifts have also been measured on individual suspended DWNTs, produced by the catalytic chemical vapor deposition (CVD) method and structurally characterized using electron microscopy, electron diffraction and Raman spectrocopy.\cite{levs11,levs15,tran17,levs17,levs17a} To our knowledge, systematic theoretical investigation of the G-band shift in DWNTs has not been reported so far, while theoretical data on this shift can be important for supporting the characterization of the DWNTs. 

The theoretical description of G-band shift has so far been hindered by computational difficulties. To begin with, the calculation of the G mode of SWNTs cannot be done accurately enough within force-constant models or models using empirical potentials, because they do not describe sufficiently well the electronic response to the atomic displacements.\cite{sait98,popo00} The latter response can be accounted for explicitly in full electronic calculations within the ab-initio approach.\cite{sanc99} A major drawback of the ab-initio models is that they become computationally very expensive with the increase of the number of atoms in the unit cell of the nanotube and cannot encompass the majority of the observable nanotube types. Alternatively, with smaller but still sufficient accuracy, the G mode can be calculated within the symmetry-adapted ab-initio-based non-orthogonal tight-binding (NTB) model.\cite{popo04} Secondly, with a few exceptions, the DWNTs do not have translational periodicity and, therefore, the all-electron models for periodic structures are not applicable. The lack of translational periodicity poses a serious problem for the estimation of the G modes, which requires special theoretical treatment. Here, we propose a computational scheme that uses the NTB model and relies on approximations for deriving the dependence of the G-band shift on the inner-layer radius and interlayer separation. We focus on the G bands of the inner layers of DWNTs, because these layers are mostly perfect and the shifts are predominantly due to interlayer interactions, contrary to the outer layers, which are influenced by the environment and often have adsorbed atoms. We constrain ourselves to semiconducting inner layers and leave out the case of metallic layers, where additional, computationally expensive corrections to the G mode, due to the strong electron-phonon interactions, are mandatory.\cite{pisc07,popo10a}       

The paper is organized as follows. The theoretical background is presented in Sec. II. The accomplished work is given and discussed in Sec. III. The paper ends up with conclusions, Sec. IV. 

\section{Theoretical background}

A SWNT can be considered as obtained by cutting out a rectangle of graphene, defined by the pair of orthogonal lattice vectors $\vec T$ and $\vec C$, and rolling the rectangle along $\vec C$ into a seamless cylinder. This rolled-up nanotube can be characterized by the radius $R = \| \vec{C} \| /2\pi$, translation period $\| \vec{T} \|$, as well as the chiral angle $\theta$, defined as the angle between $\vec C$ and the nearest zigzag of carbon atoms. All structural parameters of the rolled-up nanotube can be expressed by means of the nearest-neighbor interatomic distance and the indices $(n,m)$ of $\vec C$. Therefore, the indices $(n,m)$ specify uniquely the SWNT. Normally, the total energy of the rolled-up nanotube is not minimal and the atomic structure of the nanotube has to be subjected to relaxation in order to find the structure with minimum energy, which is a necessary step before phonon calculations. 

Furthermore, a DWNT is composed of two coaxially nested SWNTs and can be labeled as $(n_i,m_i)@(n_o,m_o)$, where the indices $i$ and $o$ denote ``inner'' and ``outer'' layer, respectively. In the general case, the two layers of a DWNT are incommensurate. The electron and phonon eigenvalue problems for such DWNTs cannot be solved by the usual computational approaches for systems with translational symmetry and one has to resort to approximations. Previously, for the structural relaxation and the calculation of the RBLMs of DWNTs, the layers were considered as elastic continuum cylinders, interacting with each other via Lennard-Jones (LJ) potential.\cite{popo02,pfei04} Alternatively, in the case of commensurate layers, the atomic structure of the layers was taken into account and minimization of the total energy, consisting of the  energy of the layers within the force-constant model and the interlayer interaction energy via LJ potentials, was carried out.\cite{popo02} We extend the latter approach to the general case of incommensurate layers by relaxing the total energy of the DWNT, expressed as the sum of total energy of each layer per unit length, derived within the NTB model, and the interlayer interaction energy per unit length, averaged over a very long piece of the DWNT.

While, in the structural relaxation step, the incommensurability problem can be overcome by the proposed approximations, this problem cannot be solved as easily for the calculation of the G mode. The straightforward approach for derivation of the shift of the G mode would be to use quantum-mechanical perturbation theory. Here, we follow a less rigorous approach, performed in two steps. First, the DWNTs are fully relaxed and the G-band shift is calculated for the relaxed structure without interlayer interactions. This shift will be referred to as the \textit{relaxation-induced shift}. For the calculation of this shift, additional external radial forces are necessary for keeping the separate layers in equilibrium. Secondly, the additional shift due to interlayer interactions between the relaxed layers is a small correction and can be estimated by perturbation theory. This shift will be referred to as the \textit{interlayer interaction-induced shift}. The use of perturbation theory for estimation of this shift is computationally expensive and we take this shift over from calculations on Bernal bilayer graphene. 

The straightforward calculation of the electronic band structure and phonons for a large variety of SWNTs is accompanied with insurmountable computational difficulties because of the very large translational unit cells of most of the SWNTs. Fortunately, the SWNTs have screw symmetry that allows reducing the computational efforts by resorting to two-atom unit cells. This symmetry-adapted approach has been used for calculation of the electronic structure\cite{popo04} and phonon dispersion\cite{popo06} of several hundred SWNTs within the NTB model. In this model, the Hamiltonian and overlap matrix elements are derived as a function of interatomic separation from an ab-initio study on carbon dimers\cite{pore95} and the Slater-Koster scheme is adopted for the angular dependence of the matrix elements.

\section{Results and Discussion}

\subsection{Relaxed DWNT structure}

\begin{figure}[tbph]
\includegraphics[width=80mm]{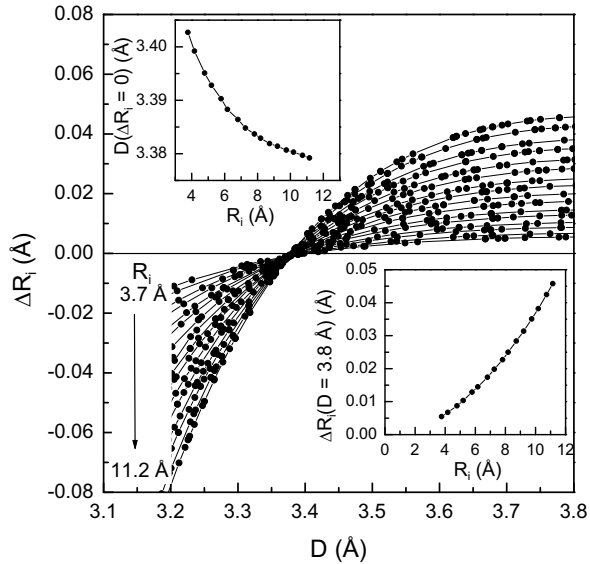} 
\caption{Relaxation-induced change of the inner layer radius of DWNTs, $\Delta R_i$, for DWNTs with given $R_i$ from $3.7$ to $11.2$ \AA~ and different $R_o$. The upper inset shows $D_0$=$D(\Delta R_i=0)$ vs $R_i$.  The lower inset shows the dependence of $\Delta R_i$ on $R_i$ at $D = 3.8$ \AA. The calculated data are drawn by solid circles. The lines are guides for the eye.}
\end{figure}

We consider DWNTs with semiconducting inner layers with radii in the interval from $3.7$ to $11.2$ \AA~ with a step of about $0.5$  \AA, namely, $(6,5)$, $(8,4)$, $(8,6)$, $(10,5)$, $(13,3)$, $(14,3)$, $(11,9)$, $(18,1)$, $(12,11)$, $(14,10)$, $(22,4)$, $(16,11)$, $(22,5)$, $(16,14)$, $(21,10)$, and $(25,6)$ in order of increasing radius. The outer layers are all layers for which the unrelaxed interlayer separation falls in the interval between $3$ and $4$ \AA~and which corresponds to the observable interlayer separations. The structural relaxation of the DWNTs is performed within the NTB model.\cite{popo04} The circular cross-section and the coaxiality of the layers is preserved during the relaxation procedure. For calculating the average interlayer interaction energy, a $100$ \AA-long piece of the DWNTs is considered, for which the average interlayer energy converges below $10^{-7}$ eV. The relaxed radii of the isolated layers will be denoted as $R_{i0}$ and $R_{o0}$, while those of the relaxed layers of the DWNTs will be denoted as $R_i$ and $R_o$.

In Fig. 1, the calculated change of the inner layer radius $\Delta R_i = R_i - R_{i0}$ of the relaxed DWNTs is shown for the considered inner layers as a function of the relaxed interlayer separation $D = R_o - R_i$ for various outer layers. It can be seen, that with increasing separation, $\Delta R_i$ increases from negative to positive values, changing sign close to $D \approx 3.4$ \AA. This behavior can be explained with the pressure on the layers due to the interlayer interactions. For $D < 3.4$ \AA, the pressure is ``positive'' and tends to shrink the inner layer, while for $D > 3.4$ \AA, the pressure is ``negative'' and expands the inner layer.\cite{cris07,liu13,levs15} 

The curves for a given inner layer and different outer layers cross the horizontal line $\Delta R_i = 0$ at $D \approx 3.4$ \AA. The separation $D_0$=$D(\Delta R_i=0)$ vs $R_i$ decreases exponentially from $3.40$ to $3.38$ \AA~for $R_i$ increasing from $3.7$ to $11.2$ \AA~(Fig. 1, upper inset). In the limiting case of $R_i \rightarrow \infty$, $D_0$ should tend to that for graphite of $\approx 3.35$~\AA.

\begin{figure}[tbph]
\includegraphics[width=80mm]{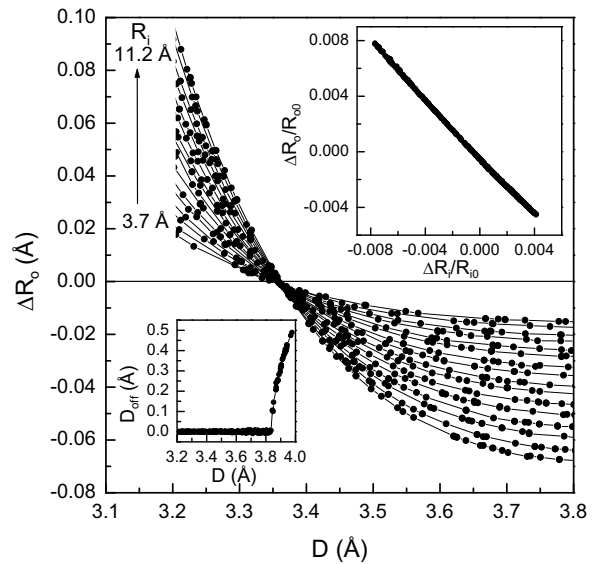} 
\caption{Relaxation-induced change of the outer layer radius of DWNTs, $\Delta R_o$, for DWNTs with given $R_i$ from $3.7$ to $11.2$ \AA~ and different $R_o$. The upper inset shows the linear dependence of $\Delta R_o/R_{o0}$ vs $\Delta R_i/R_{i0}$. The lower inset shows the distance between the axes of the layers, $D_{off}$ vs $D$. The calculated data are drawn by solid circles. The lines are guides for the eye.}
\end{figure}

At a given $D$, the absolute value of  $\Delta R_i$ increases with increasing  $R_i$  as a second-degree power function. This is illustrated for $D = 3.8$ \AA~ in the lower inset of Fig. 1.

In Fig. 2, the results for the change of the outer radius $\Delta R_o = R_o - R_{o0}$ are presented in a manner, similar to these for $\Delta R_i$. The upper inset in Fig. 2 shows the dependence of $\Delta R_o/R_{o0}$ on $\Delta R_i/R_{i0}$. The almost equal relative changes of the radii for the inner and outer layers can be explained by a simple mechanical model. Each isolated layer has a RBM with frequency $\omega = \sqrt{\kappa \slash m}$, where $m$ and $\kappa$ are layer's mass and spring constant, respectively. The changes of the equilibrium radii of the two layers of a DWNT, $\Delta R_o$ and $\Delta R_i$, due to switching-on of coupling between them, satisfy the relation $\kappa_o \Delta R_o=- \kappa_i \Delta R_i$. Bearing in mind that $\omega \propto 1/R$ and $m \propto R$, where the proportionality coefficients are equal for both layers,\cite{popo06} we arrive at the relation $\Delta R_o/R_{o0}=-\Delta R_i/R_{i0}$. 
 
The calculations show that for interlayer separations $D > 3.8$ \AA~ the DWNT structure becomes unstable with respect to off-axial displacement of the inner layer with respect to the outer layer and the equilibrium structure is characterized by nonzero distance between the axes of the two layers, $D_{off}$,\cite{pfei04} (Fig.2, lower inset). Such off-axial configurations of DWNTs have not been observed yet, but the proposed computational scheme can be extended to encompass such cases as well.

\begin{figure}[tbph]
\includegraphics[width=80mm]{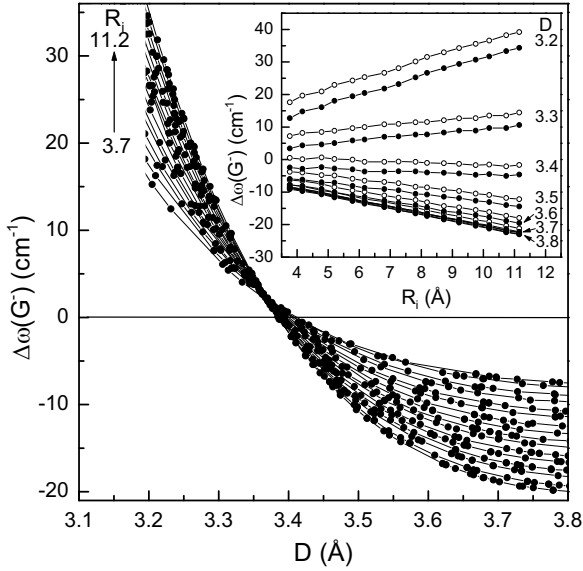} 
\caption{Relaxation-induced shift of the G$^-$ band of the inner layer  for DWNTs with given $R_i$ from $3.7$ to $11.2$ \AA~ and different $R_o$ (solid circles). Inset: interpolation-derived relaxation-induced shift vs $R_{i}$ at given values of $D$ (open circles) and total shift (solid circles). The lines are guides for the eye.}
\end{figure}

\subsection{G-band shift}

\begin{figure}[tbph]
\includegraphics[width=80mm]{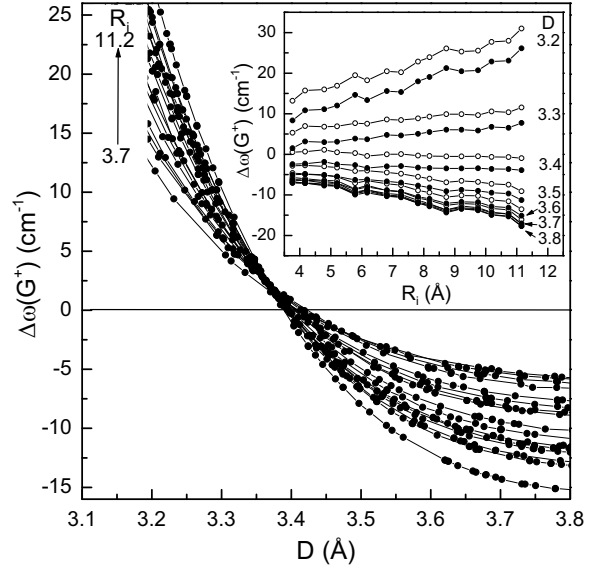} 
\caption{Relaxation-induced shift of the G$^+$ band of the inner layer  for DWNTs with given $R_i$ from $3.7$ to $11.2$  \AA~ and different $R_o$ (solid circles). Inset: interpolation-derived relaxation-induced shift $R_{i}$ at given values of $D$ (open circles) and total shift (solid circles). The lines are guides for the eye.}
\end{figure}

The \textit{relaxation-induced shifts} of the G$^-$ and G$^+$ bands of the inner layer of the relaxed DWNTs with respect to the bands of the relaxed isolated layers are shown as a function of D in Figs. 3 and 4. Both figures exhibit a general trend of decreasing of the shift, changing the sign of the shift from positive to negative at $\approx 3.4$ \AA, and further increasing the shifts in absolute value with increasing $D$. There is a correlation of this behavior with that of $\Delta R_i$ vs $D$. Namely, for $D < 3.4$ \AA, the G bands are blue-shifted due to decreased interatomic distances in the inner layer, while for $D > 3.4$ \AA, the G bands are red-shifted due to increased interatomic distances in the inner layer. Both bands have almost zero shift close to $D = 3.4$ \AA. The G-bands shifts of the outer layers undergo opposite changes vs D (not shown). For a given DWNT, in absolute value, the G$^-$-band shift is larger than the G$^+$-band shift. The G$^-$- and G$^+$-band shifts for the inner layer fall in the range between $-20$ and $35$ cm$^{-1}$, and $-15$ and $25$ cm$^{-1}$, respectively. For a given $D$, the G-band shifts increase in absolute value with increasing $R_i$ (insets of Figs. 3 and 4, open circles). The dependence of the shifts on $R_i$ for a given $D$ is almost linear with minor deviations from linearity for the G$^+$-band shift.

The \textit{interaction-induced shifts} of the G bands should be treated in perturbation theory. This approach can be extremely computer-time consuming because it has to be accomplished at the quantum-mechanical level. Here, we follow an alternative route and calculate exactly the effect of interlayer interactions on the G mode (Raman-active E$_{2g}$ mode) in Bernal bilayer graphene. For this purpose, the bilayer structure is relaxed at fixed interlayer separations and the G mode is calculated with and without interlayer interactions using the NTB model. The switching-on of the interlayer interactions lifts the degeneracy of the two E$_{2g}$ Raman-active modes of the two layers and gives rise to two modes of E$_{2g}$ and E$_{1u}$ symmetry, shifted downwards and upwards, respectively. We are interested in the modification of the former, which is downshifted by the interlayer interactions. In particular, with increasing the layer separation from $D = 3.2$ to $3.8$ \AA~ with a step of $0.1$ \AA, the G-mode shift decreases in absolute value and has the following values: $-4.87$, $-3.81$, $-2.96$, $-2.19$, $-1.56$, $-1.06$, and $-0.67$ cm$^{-1}$. These values are comparable to the ab-initio derived ones and agree well with the available experimental data.\cite{sun15} The obtained results for bilayer graphene can be transferred to DWNTs, because of the similar local environment of the atoms of the two layers. Namely, an atom of a given layer interacts with a small, almost planar portion of the other layer. For small radius of the layers, small deviations from the results for bilayer graphene, due to curvature effects, can be expected.

\begin{figure}[tbph]
\includegraphics[width=80mm]{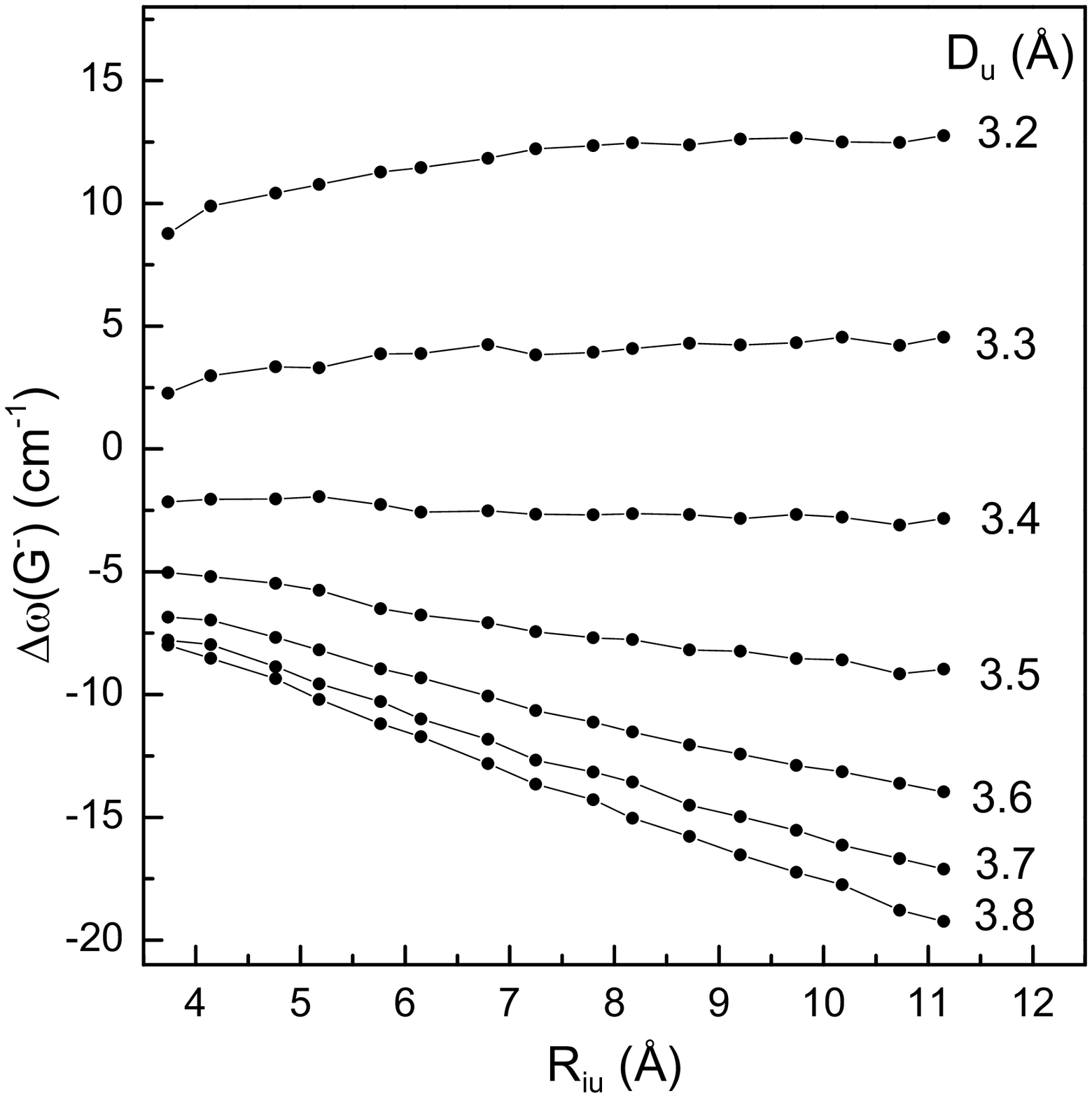} 
\caption{Total shift (solid circles) of the G$^-$ band of the inner layer of DWNTs as a function of $R_{iu}$ for different $D_u$.  The lines are guides for the eye.}
\end{figure}

Finally, the total G-band shift can be written as the sum of the relaxation-induced shift and the interlayer interaction-induced one, where the former is calculated for the particular relaxed DWNT with switched-off interlayer interactions and the latter is taken over from bilayer graphene. The resulting total shifts are presented as a function of the relaxed inner layer radius $R_{i}$ for several values of the interlayer separation D in the insets of Figs. 3 and 4 (solid circles). For a given $D$, the total G-band shifts increase in absolute value with increasing $R_i$. The resulting curves of these dependencies are relatively smooth for the G$^{-}$ band but have wiggles in the case of the G$^{+}$ band. The latter can be explained with the dependence of this mode not only on the radius but also on the translation period, because of the longitudinal character of this mode.

\subsection{Comparison to experiment}

\begin{figure}[tbph]
\includegraphics[width=80mm]{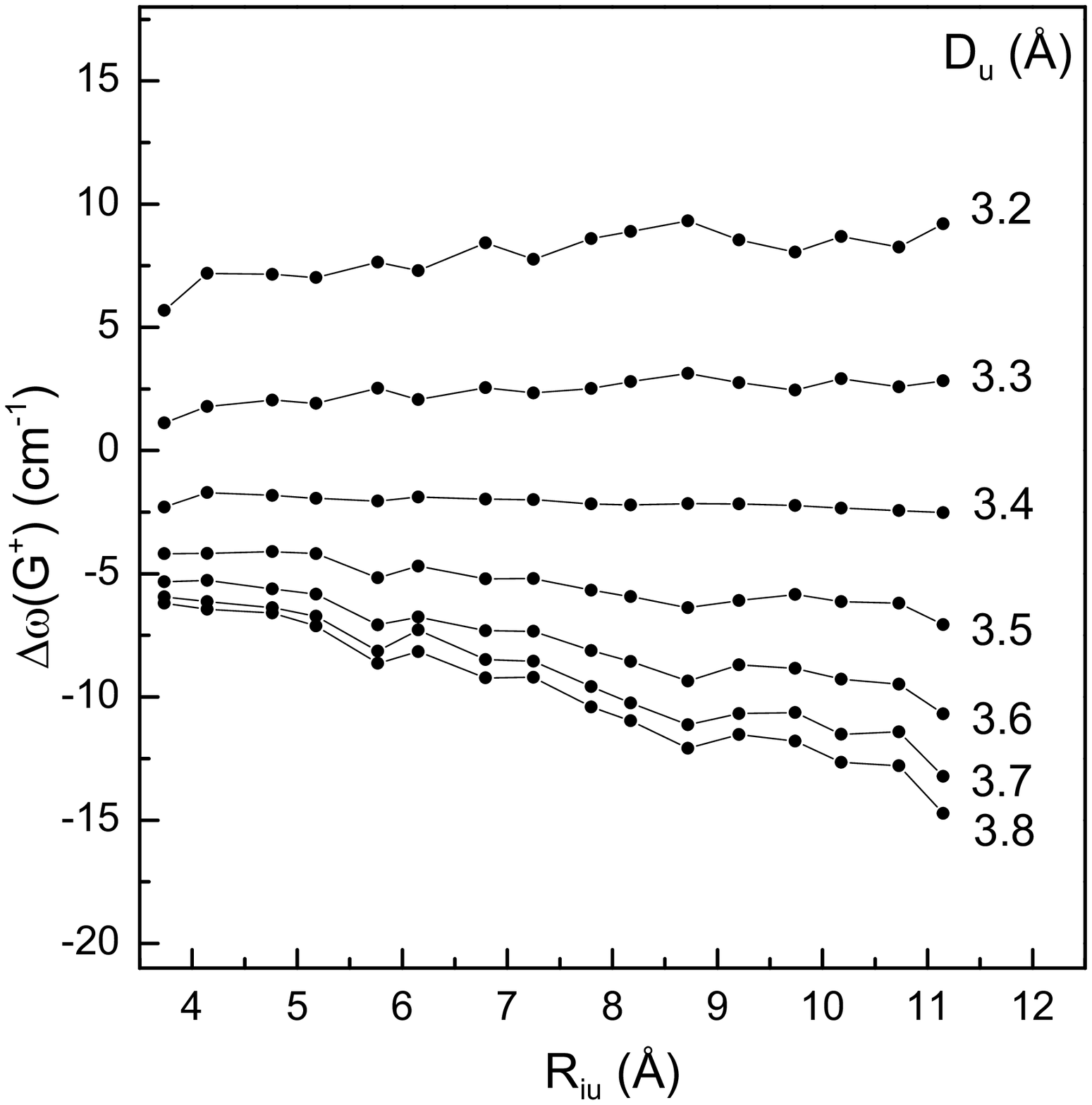} 
\caption{Total shift (solid circles) of the G$^+$ band of the inner layer of DWNTs as a function of $R_{iu}$ for different $D_u$.  The lines are guides for the eye.}
\end{figure}

The relaxed layer radii and the corresponding interlayer separation of DWNTs are difficult to determine with sufficient accuracy from experiment (e.g. electron diffraction). Hence, it is a common practice in the literature to use unrelaxed isolated layers radii $R_{iu}$ and $R_{ou}$, which are calculated directly from ($n_{i(o)}$,$m_{i(o)}$) by the usual relation $R_{i(o)u}=\sqrt{3}a_{C-C}\sqrt{m_{i(o)}^2+m_{i(o)}n_{i(o)}+n_{i(o)}^2} \slash 2\pi$ with $a_{C-C}$=1.42 \AA. The unrelaxed interlayer distance is defined as $D_u =  R_{ou} - R_{iu}$. For the practical purposes of the comparison with experimental data, we now plot, in Figs. 5 and 6, the obtained results for the total calculated G$^-$-band shifts and the total calculated G$^+$-band shifts, respectively, as a function of $R_{iu}$ and for several values of $D_u$. These plots permit to evaluate the shift of the G$^{-}$ and G$^{+}$ bands of the inner layer for given $D_u$ and $R_{iu}$.\\

In Fig. 7, we compare our theoretical data for the  G$^-$-band shifts (solid circles) with experimental ones obtained (i) on several $C_{60}$-derived DWNTs\cite{villa10} (green open triangles),  and (ii) on individual suspended index-identified CVD-DWNTs\cite{levs15,levs17a} (red open circles). Fig. 8 displays the comparison between the calculated G$^+$-band shifts (solid circles) and experimental G$^+$-band shifts (red open circles) measured on individual suspended index-identified CVD-DWNTs.\cite{levs15,levs17a}\\

Concerning the results obtained on $C_{60}$-derived DWNTs, all tubes are $(6,5)@(n_o,m_o)$. The experimental G$^-$-band shifts are directly calculated from the frequencies of the G$^-$ band, given in Ref.\cite{villa10} versus the one measured on the (6,5) SWNT.\cite{telg12} The index assignment of the outer tubes of these DWNTs, and hence the $D_u$ interlayer distances, was revised using a more accurate approach (coupled-oscillator model \cite{liu13}) than in Ref. \cite{villa10} (for details, see Ref.\cite{tran17}). The data are given in Table I.\\

\begin{figure}[tbph]
\includegraphics[width=80mm]{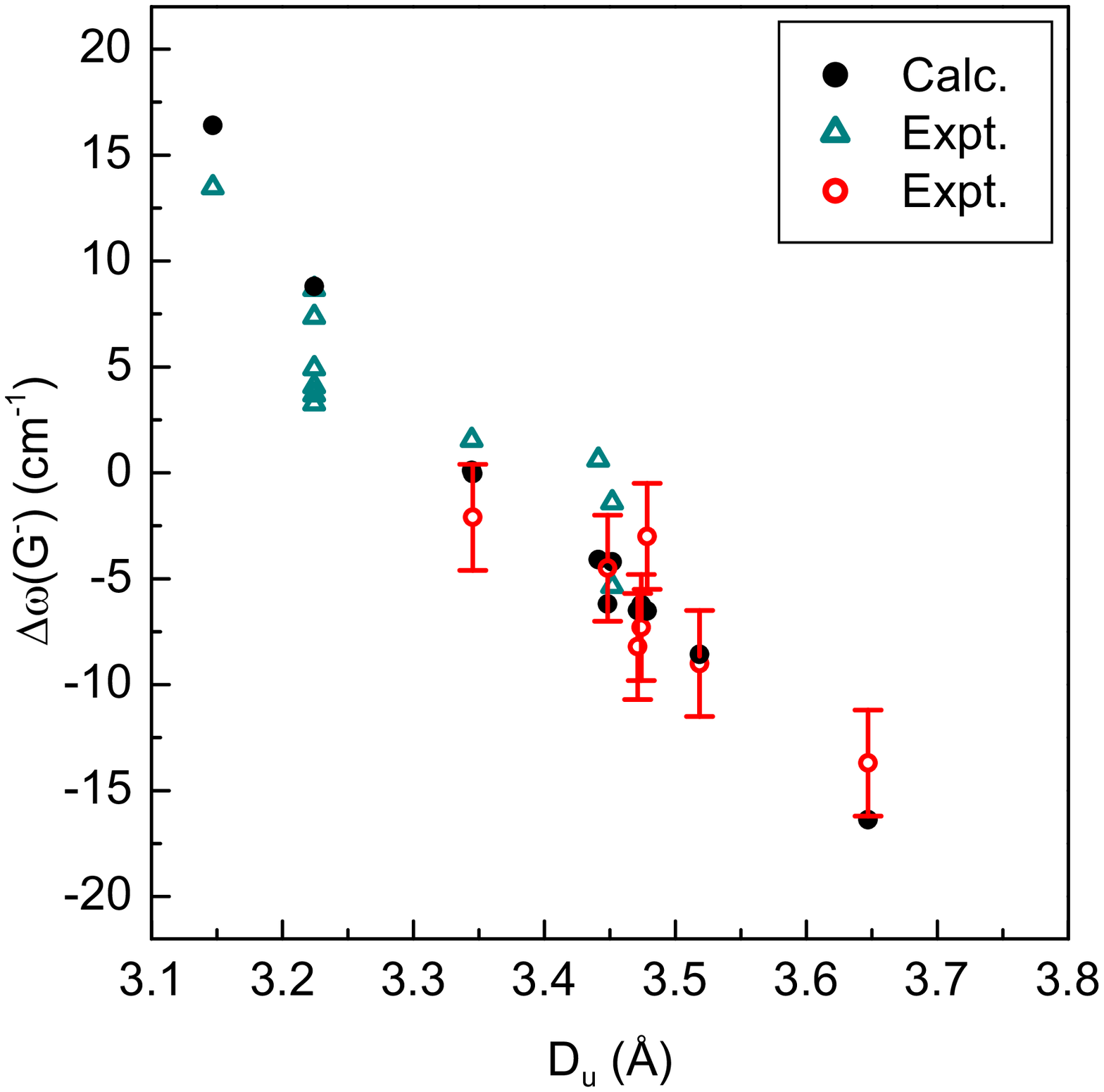} 
\colorcaption{Comparison of calculated (solid circles) with  experimental G$^-$-band shifts for several $C_{60}$-derived DWNTs\cite{villa10} (green open triangles) and  for several index-identified DWNTs (red open circles) (Ref.\cite{levs15,levs17a} and this work).}
\end{figure}

\begin{table}[b]
\caption{\label{tab:table1}
Comparison of experimental G$^{-}$-band shifts\cite{villa10} and calculated ones  in cm$^{-1}$ for several $C_{60}$-derived $(6,5)@(n_o,m_o)$ DWNTs. In comparison with Ref.\cite{villa10} the $(n_o,m_o)$ used here were re-attributed accounting for the interlayer interactions between the layers within a coupled-oscillator model (without structural relaxation, see text).\cite{tran17}
}
\begin{ruledtabular}
\begin{tabular}{ccrr}
&&\multicolumn{2}{c}{$\Delta \omega (G^-)$}\\
\textrm{$(n_o,m_o)$\cite{tran17}}&
\textrm{$D_u$ (\AA)\cite{tran17}}&
\textrm{Expt.\cite{villa10}} &
\textrm{Calc.}\\
\colrule
$(13,8)$ & $3.45$ & $-1.4$ & $-4.2$ \\
$(13,8)$ & $3.45$ & $-5.3$ & $-4.2$ \\
$(16,4)$ & $3.44$ & $-0.6$ & $-4.1$ \\
$(17,2)$ & $3.35$ & $-1.6$ & $0.1$ \\
$(14,6)$ & $3.22$ & $3.4$ & $8.8$ \\
$(14,6)$ & $3.22$ & $7.4$ & $8.8$ \\
$(14,6)$ & $3.22$ & $3.7$ & $8.8$ \\
$(14,6)$ & $3.22$ & $4.9$ & $8.8$ \\
$(14,6)$ & $3.22$ & $8.7$ & $8.8$ \\
$(14,6)$ & $3.22$ & $4.1$ & $8.8$ \\
$(13,7)$ & $3.15$ & $13.5$ & $16.4$ \\
\end{tabular}
\end{ruledtabular}
\end{table}

Concerning the individual suspended index-identified CVD-DWNTs, the experimental data come from Ref.\cite{levs15,levs17a} and from this work (these additional DWNTs are indicated in Table II). The experimental shifts are estimated relative to the G-band frequencies of isolated layers (SWNTs). For the G$^-$ band, the SWNT reference frequencies were calculated as the average of the values obtained using an empirical law\cite{telg12} and a theoretical law\cite{pisc07}. In the case of the G$^+$ band, the experimental shifts are estimated relative to the G-band frequencies of isolated layers, calculated as the average of the values of an empirical law\cite{telg12} and 1592 cm$^{-1}$ (Ref.\cite{pail06}). The errors on the estimated shifts are set at $2.5$ cm$^{-1}$, by summing the experimental uncertainty of $1$ cm$^{-1}$ on the measured DWNTs G-band frequencies and the maximum difference ($1.5$ cm$^{-1}$) between the SWNTs G-band frequencies values deduced using the different laws. The experimental and calculated values of the shifts are given in Table II.\\ 

\begin{figure}[tbph]
\includegraphics[width=80mm]{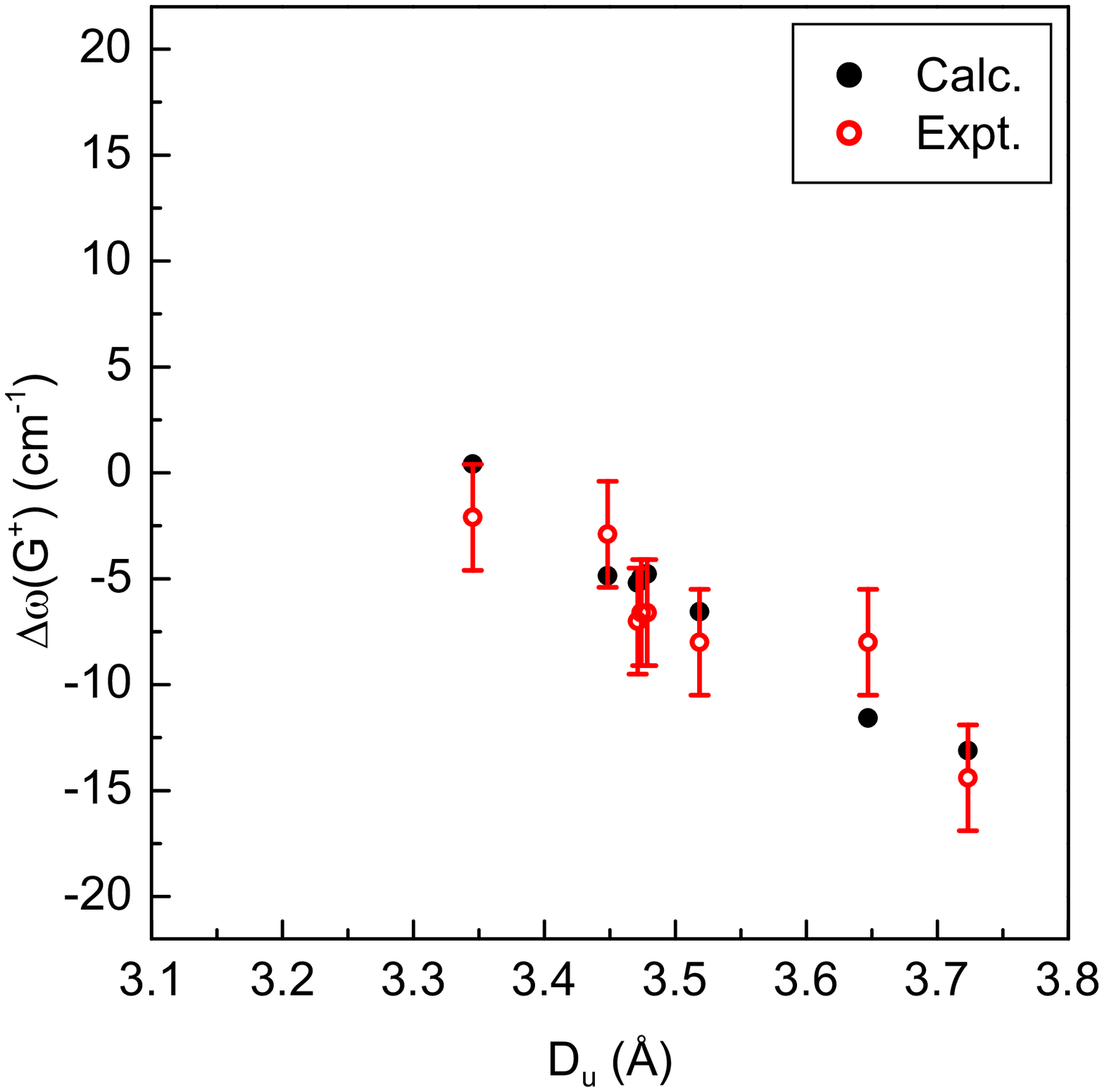} 
\colorcaption{Comparison of calculated (solid circles) with experimental (red open circles) G$^+$-band shifts for several index-identified DWNTs (Ref.\cite{levs15,levs17a} and this work).}
\end{figure}

\begin{table}[b]
\caption{\label{tab:table2}
Experimental\cite{levs15,levs17a}$^{,}$\footnote{This work.} and calculated G-band shifts  in cm$^{-1}$ for several index-identified DWNTs.  Empirical and theoretical relations are used for estimation of the G-band frequencies of non-interacting layers (see text)). 
}
\begin{ruledtabular}
\begin{tabular}{lccrrrr}
&&&\multicolumn{2}{c}{$\Delta \omega (G^+)$}&\multicolumn{2}{c}{$\Delta \omega (G^-)$}\\
\textrm{$(n_i,m_i)@(n_o,m_o)$}&
\textrm{$D_u$ (\AA)}&\textrm{$D$ (\AA)}&
\textrm{Expt.}&\textrm{Calc.}&\textrm{Expt.}&\textrm{Calc.}\\
\colrule
$(12,8)@(16,14)$   & $3.35$ & $3.35$ & $-1.6$ & $0.4$ & $-2.1$ & $0.0$ \\
$(16,12)@(27,10)$ & $3.45$ & $3.42$ & $-2.4$ & $-4.9$ & $-4.5$ & $-6.2$ \\
$(14,7)@(21,10)^a$   & $3.48$ & $3.45$ & $-6.5$ & $-5.2$ & $-8.2$ & $-6.5$ \\
$(10,9)@(18,11)$   & $3.48$ & $3.45$ & $-6.1$ & $-4.9$ & $-7.3$ & $-6.2$ \\
$(16,2)@(16,14)^a$   & $3.49$ & $3.46$ & $-6.1$ & $-4.8$ & $-3.0$ & $-6.5$ \\
$(13,9)@(24,7)$     & $3.52$ & $3.49$ & $-7.5$ & $-6.6$ & $-9.0$ & $-8.6$ \\
$(22,11)@(27,17)$ & $3.65$ & $3.56$ & $-7.5$ & $-11.6$ & $-13.7$ & $-16.4$ \\
$(30,1)@(27,19)^a$   & $3.73$ & $3.62$  & $-14.4$ &$-13.1$ & $-$ &   $-19.3$\\
\end{tabular}
\end{ruledtabular}
\end{table}

Overall, the plotted theoretical data for the G-band shift follow rather well the trend of the experimental ones (Figs. 7 and 8). For the $C_{60}$-derived DWNTs, the deviation  of the G$^{-}$-band shift and the dispersion of the results could be due to the influence of the the substrate which is not included in this study. 
For the suspended DWNTs, the interlayer separations $D_u$ range from $3.35$ to $3.73$ \AA. The DWNT with $D_u$ close to $3.35$ \AA~is almost undeformed by the interlayer interactions, while that with $D_u$ close to $3.73$ \AA~undergoes radius change as large as about $0.11$ \AA. The predicted shifts of both inner-layer G bands for the former one are almost zero and the predicted shifts for the latter one are $-13.1$ and $-19.3$ cm$^{-1}$. All calculated shifts agree well with the measured ones within the experimental uncertainty except for the two DWNTs with largest $D_u$. The origin of the deviations from the measured values can be sought in the existence of factors, which are not accounted for in the current calculations. For example, due to the large radius of the latter two DWNTs, possible non-circular deformations of the layers could modify the G modes and produce smaller G band shifts. More sophisticated calculations should be performed in order to throw light on the origin of this discrepancy and yield better agreement with experiment.\\

\section{Conclusions}

We have presented calculations of the G-band shift of DWNTs within a non-orthogonal tight-binding model. The lack of translational periodicity of most of the observed DWNTs predetermines the use of approximations, namely, the shift is calculated in two feasible steps. First, the DWNTs structure is relaxed and the shift is calculated for the relaxed structure but for switched-off interlayer interactions. Secondly, contribution of the interlayer interactions, derived for bilayer graphene, is added to the shift. The agreement with experiment is excellent in most cases, but certain discrepancy is observed in a few cases. The latter could be resolved by future elaborations of the computational approach.

\acknowledgments

VNP and DL acknowledge visitor grants at L2C Montpellier from CNRS. DL acknowledges Metchnikov grant from the French Embassy in Russia. JLS and MP acknowledge financial support by the ANR GAMBIT project, grant ANR-13-BS10-0014 of the French Agence Nationale de la Recherche.


%

\end{document}